\documentstyle[preprint,aps,epsfig]{revtex}
\begin{document}
\draft
\title{QCD Sum Rules for the Isospin-Breaking Axial Correlator\\
with Correct Chiral Behavior}
\author{Kim Maltman} 
\address{Department of Mathematics and Statistics, York University, \\
          4700 Keele St., North York, Ontario, CANADA M3J 1P3}
\address{Special Research Center for the Subatomic Structure of Matter, \\
          University of Adelaide, Australia 5005}
\author{Michael C. Birse}
\address{Theoretical Physics Group, Department of Physics and Astronomy, \\
University of Manchester, M13 9PL, UK}
\maketitle
\begin{abstract}
We revisit the QCD sum-rule treatment of the isospin-breaking correlator
$\langle 0|T[A_\mu^3(x) A_\nu^8(0)]|0\rangle$, in light of the recent claim
that a previous treatment produced results incompatible with known chiral
constraints. The source of the error in the previous analysis is identified,
and a corrected version of the sum-rule treatment obtained.  It is then shown
that, using input from chiral perturbation theory, one may use the resulting
sum rule to extract information on the leading chiral behavior of
isospin-breaking parameters associated with the coupling of excited
pseudoscalar resonances to the axial currents.  A rather accurate extraction
is possible for the case of the $\eta^\prime$. Demanding stability of the
sum-rule analysis also allows us to improve the upper bound on the fourth-order
low-energy constant, $L_7$.
\end{abstract}

\section{Introduction}
 
One of the most attractive features of chiral perturbation theory
(ChPT)\cite{gl85} is that it provides a framework for constructing effective
hadronic Lagrangians in the most general possible way that implements both the
symmetries of QCD and the symmetry breaking pattern of the approximate chiral
symmetries of QCD. As such, it fully incorporates the consequences of QCD in
the low-energy regime. The price to be paid for using only symmetry arguments
is that every term in the effective Lagrangian, {${\cal L}_{\rm eff}$},
allowed by these arguments will appear, multiplied by an undetermined constant
(referred to as a low-energy constant, or LEC).  These LEC's could, in
principle, be computed from QCD, but must be treated as parameters to be
determined phenomenologically if one does not go beyond ChPT.

Although such effective Lagrangians are necessarily non-renormalizable,
Weinberg's counting argument\cite{weinberg79} shows that only a finite number
of terms in ${\cal L}_{\rm eff}$ contribute if one expands to fixed ``chiral''
order, that is in powers of the external momenta (generically denoted as $p$)
and current quark masses (where $m_q$ counts as order ${\cal O}(p^2)$). As a
result, in the chiral expansion of any low-energy observable, the general form
of the dependence on external momenta and light quark masses, to a given
order, can be computed straightforwardly from the form of the relevant terms
in ${\cal L}_{\rm eff}$.

Since this formal dependence is a rigorous consequence of the symmetries and
approximate symmetries of QCD, it follows that ChPT can be used to place
constraints on treatments of the same observable using other methods.  Indeed,
if one makes a chiral expansion of the results obtained by any other method
and finds that terms present in ChPT to a given order are missing, then one
knows unambiguously that either the method itself, or some truncation employed
in it, is incompatible with QCD.  This is true regardless of the rapidity of
convergence of the chiral series in question:  all terms required by the
symmetries of QCD must be present if the method is to correctly incorporate
the consequences of QCD.

An example of the use of such constraints is provided by 
the analysis of the nucleon mass using QCD sum rules.
Standard treatments were shown to produce an
expression for $m_N$ in terms of condensates that implies the presence of
certain chiral logarithms in $m_N$, although such contributions are known from
ChPT to be absent\cite{gk94,lccg95}. The source of this problem was found to be
a failure to treat properly the contribution of the $\pi N$ continuum to the
sum rule in the original analyses\cite{lccg95}; including the leading
contributions from such states restores the correct chiral behavior of $m_N$.

A more severe problem of the same type has been pointed out in the case of
the isospin-breaking axial correlator
\begin{equation}
\Pi_{\mu\nu}^{38}(q)= i\int d^4x\ e^{iq.x}
\langle 0\vert T(A^3_\mu (x)A^8_\nu (0))\vert 0\rangle
\equiv \Pi_1(q^2)q_\mu q_\nu - \Pi_2(q^2)q^2g_{\mu\nu} \label{pidefn}\ ,
\end{equation}
where $A^3_\mu$, $A^8_\nu$ are the neutral members of the octet of axial
currents $A^a_\mu =\bar q\gamma_\mu\gamma_5{\lambda^a\over 2}q$. This
correlator was first analyzed using QCD sum rules in Ref.~\cite{chm} (CHM).
As shown in Ref.~\cite{km95}, however, if one writes $\Pi_1(q^2)$,
which contains the $\pi^0$ and $\eta$ pole contributions, in the form
\begin{equation}
\Pi_1(q^2)=\left( {\frac{g_\eta}{ q^2-m_\eta^2}}
-{\frac{g_\pi}{ q^2-m_\pi^2}}\right)+\cdots
\ =\left( {\frac {q^2(g_\eta -g_\pi )+(g_\pi m_\eta^2 -g_\eta m_\pi^2)}
{(q^2 -m_\eta^2)(q^2-m_\pi^2)}}\right)+\cdots \ ,\label{poleterms}
\end{equation}
then the expression for $g_\eta -g_\pi$ (given by the slope of the numerator
with respect to $q^2$) obtained from the sum-rule analysis is lacking both the
leading analytic and leading non-analytic terms from its chiral expansion in
terms of the light quark masses.  (The demonstration of this is reviewed
briefly below in Sec.~II.)

In this paper we revisit the sum-rule analysis of the axial correlator above,
and identify the source of this problem. We then obtain a corrected version of
the relevant sum rule and show how it can be used to extract information on
isospin-breaking couplings of the higher pseudoscalar resonances.

The paper is organized as follows.  In Sec.~II, we briefly review the sum-rule
and ChPT analyses of the correlator.  In Sec.~III, we identify the problem
with the previous sum-rule treatment and work out the corrected version of the
relevant sum rules.  In Sec.~IV, we show how one can use information from ChPT
as input into the sum rule. We also clarify the physical content of the
corrected sum rule, extracting in the process information on the
isospin-breaking couplings of the higher pseudoscalar resonances to the axial
currents.  We conclude in Sec.~V with a brief summary.

\section{Previous ChPT and sum-rule treatments}

We provide here only a very brief review, which will serve also to fix
notation.  For more details the reader is referred to Refs.~\cite{chm}
and \cite{km95} for the sum-rule and ChPT treatments, respectively.

We first review the sum-rule treatment.\footnote{See, for example,
Refs.\cite{svz1,svz2,RRY85,narisonbk} for details of the general method of QCD
sum rules.} As usual, the aim is to write dispersion relations for
$\Pi_1(q^2)$ and $\Pi_2(q^2)$ which relate integrals over the relevant
physical spectral functions to the behaviors at large spacelike $q^2$, where
the operator product expansion (OPE) becomes valid.  One then Borel transforms
the resulting dispersion relations in order to exponentially suppress the
higher-energy portions of the spectral integral on the phenomenological side
and simultaneously factorially suppress the contributions of higher dimension
operators on the OPE side. The scalar correlators $\Pi_1(q^2)$ and
$\Pi_2(q^2)$ in Eq.~(\ref{pidefn}) have been chosen in such a way that, from
the asymptotic behavior of $\Pi^{38}_{\mu\nu}(q)$ in QCD, it is known that the
relevant spectral integrals converge without subtraction. Note that the
definition of $\Pi_2$ employed here agrees with that used in Ref.~\cite{chm},
but differs from that in Ref.~\cite{km95} by a factor of $-q^2$.

On the phenomenological side, the axial-vector resonances contribute to both
$\Pi_1(q^2)$ and $\Pi_2(q^2)$. In the narrow-width approximation, their
contributions to the complete spectral function are written
\begin{equation}
{\frac{1}{\pi}}\left( {\rm Im}\, \Pi^{38}_{\mu\nu}(q)\right)_A
= \sum_A\, g^{(A)}\left[ -g_{\mu\nu}+q_\mu q_\nu /M_A^2\right]\, \delta
(q^2-M_A^2)\ .\label{axialspec}
\end{equation}
The pseudoscalar resonances, in contrast, contribute only to $\Pi_1$.
Following the convention of earlier works, we write these contributions
as
\begin{equation}
{\frac{1}{\pi}}\Bigl( {\rm Im}\, \Pi_1(q^2)\Bigr)_P = g_\pi \delta (q^2
-m_\pi^2)-g_\eta \delta (q^2 -m_\eta^2)+g_{\eta^\prime}\delta (q^2
- m_{\eta^\prime}^2)+g_{\pi^\prime}\delta (q^2-m_{\pi^\prime}^2)+\cdots
\ .\label{spectral}
\end{equation}
(The minus sign in front of $g_\eta$ is conventional and related to the
fact that, so defined, $g_\eta =g_\pi$ at leading order in the chiral
expansion.)

On the OPE side of the sum rule, the expressions for the scalar correlators
have been worked out by CHM up to and including operators of dimension $6$,
and to order ${\cal O}(m_q,\ \alpha_s)$. Neglecting electromagnetic effects,
the results have the form (with $Q^2=-q^2$)
\begin{eqnarray}
\Pi_1(q^2) &=&{\frac{1}{4\sqrt{3}}}
\left[ C_0\ln Q^2 + {C_1\over Q^2} + {C_2\over Q^4} +
{C_3\over Q^6}\right]\ , \nonumber \\
\Pi_2(q^2) &= &{\frac{1}{4\sqrt{3}}}
\left[ C_0\ln Q^2 + {C_1\over Q^2} - {C_2\over Q^4} +
{C_3\over Q^6}\right]\ ,
\label{ope}
\end{eqnarray}
where $C_0$ and $C_2$ vanish at the level of the truncations noted above, and
\begin{eqnarray}
C_2 &=& 2[m_u\langle\bar uu\rangle - m_d\langle\bar dd\rangle]\ ,\nonumber \\
C_3 &=& {\frac{352}{81}}\pi\alpha_s[\langle\bar uu\rangle^2 - \langle\bar dd
\rangle^2] \ .
\label{opecoeff}
\end{eqnarray}
If one were to include higher-order terms in Eq.~(\ref{ope}), $C_0$ would
receive contributions at order ${\cal O}(\alpha_{EM}, \alpha_s, m_q^2)$ and
$C_1$ at ${\cal O}(m_q^2)$.  An argument analogous to that of Shifman,
Vainshtein and Zakharov\cite{svz2} for the corresponding isospin-conserving
correlator $\Pi^{33}_{\mu\nu}(q)$ shows that the higher-dimension operators
not included in these expressions are also all explicitly of order ${\cal
O}(m_q^2)$.  The form of the dimension $6$ coefficient $C_3$ in
Eq.~(\ref{opecoeff}) has been obtained assuming vacuum saturation.

As can be seen from the Lorentz structure of Eq.~(\ref{axialspec}), it is
possible to remove the contributions of the axial-vector mesons by considering
the combination
\begin{equation}
\Pi_P(q^2)\equiv \Pi_1(q^2)-\Pi_2(q^2)\ .  
\end{equation}
CHM, motivated by this observation, write a dispersion relation for
$\Pi_P(q^2)$ in the form
\begin{equation}
\Pi_P(q^2)=\int {\frac{1}{\pi}}{\frac{{\rm Im}\, \Pi_P(s)}
{s-q^2}}\, ds\ .
\label{dispreln}\end{equation}
When Borel transformed, this relation gives CHM's sum rule,
\begin{equation}
2C_2\left({\frac{1}{ M^4}}\right) = {\frac{4\sqrt3} {M^2}}\left[ g_\pi
\ e^{-{m^2_\pi /M^2}} - g_\eta \ e^{-{m^2_\eta / M^2}}\right] + \cdots\ ,
\label{chmsr}
\end{equation}
where $M$ is the Borel mass parameter and the dots refer to the contributions
of higher pseudoscalar resonances. CHM then neglect higher resonance
contributions and use this sum rule, together with its derivative
with respect to the Borel mass, $M$, to solve for $g_\eta$ and
$g_\pi$.  This 
procedure leads to their result
\begin{equation}
\left( g_\eta -g_\pi\right)_{CHM} =  
\left( {\frac{C_2}{2\sqrt3 M^2}}\right)\left[{\frac
{e^{m^2_\eta /M^2}\left( M^2+
m^2_\pi\right) - \ e^{m^2_\pi /M^2}\left( M^2+m^2_\eta\right)}{m^2_\eta - 
m^2_\pi}}\right]
\label{slopesr}
\end{equation}
for the slope of the numerator on the RHS of Eq.~(\ref{poleterms}). 

The analysis of $\Pi^{38}_{\mu\nu}(q)$ at next-to-leading (1-loop) order
in ChPT is straightforward, and follows standard methods.  We
employ throughout the notation of Gasser and Leutwyler\cite{gl85}.  The result
for $\Pi_2(q^2)$, recast so as to correspond to
the definition employed in this paper, is\cite{km95}
\begin{equation}
\Pi_2^{\rm 1-loop}(q^2)=-{\frac{B_0(m_d-m_u)}{\sqrt{3}q^2}}
\Biggl[ {\frac{3}{32\pi^2}}\left(\log (m_K^2/\mu^2)+1\right) 
-8L_5^r(\mu^2)\Biggr]\ ,\label{pi2chpt}
\end{equation}
where $B_0$ is the usual second-order LEC, related to the quark condensate in
the chiral limit, $\mu$ is the renormalization scale, and $L_5^r(\mu^2)$ is a
renormalized fourth-order LEC. Note that $\Pi_2(q^2)$ results solely from
contact terms (that is, terms in ${\cal L}_{\rm eff}$ that are quadratic in
the external axial sources). To this order, $\Pi_1(q^2)$ is saturated by the
$\pi^0$ and $\eta$ pole terms. From a similar analysis, one finds, for the
coefficients $g_\pi$ and $g_\eta$ appearing in Eq.~(\ref{spectral}),
\begin{equation}
g_\pi =f_\pi^2\epsilon_1\quad {\rm and}\quad g_\eta = f_\eta^2\epsilon_2
\end{equation}
where $F_\pi$, $f_\eta$ are the physical $\pi$, $\eta$ decay
constants and $\epsilon_1$, $\epsilon_2$ are isospin-breaking parameters
defined by 
\begin{eqnarray}
\langle 0\vert A^8_\mu\vert\pi\rangle &=& {\rm i}f_\pi\epsilon_1 q_\mu
\nonumber \\
\langle 0\vert A^3_\mu\vert\eta\rangle &=& -{\rm i}f_\eta\epsilon_2 q_\mu\ .
\label{epsvals}
\end{eqnarray}
The expressions for $f_\pi$, $f_\eta$, $\epsilon_1$ and $\epsilon_2$
valid to 1-loop order can be found in Ref.~\cite{gl85}.

The problem with the sum-rule treatment is exposed when one uses the
known chiral expansions of the meson masses and quark condensates to
rewrite the sum-rule result, Eq.~(\ref{slopesr}), as
\begin{equation}
\left( g_\eta -g_\pi\right)_{\rm CHM} 
=\theta_0 F^2\left( {8\over 9}{B_0^2(m_s-{\hat m})
(m_s+2{\hat m})\over M^4} +\cdots \right)\ ,\label{srch}
\end{equation}
to order ${\cal O}(m_q^2)$. Here $F$ is a second-order LEC, equal to $f_\pi$
in the chiral limit, and $\theta_0$ is the leading-order $\pi^0$--$\eta$
mixing angle,
\begin{equation}
\theta_0 ={\frac{\sqrt{3}}{4}}\left( {\frac{m_d-m_u}{m_s-\hat{m}}}\right)\ ,
\end{equation}
with $\hat{m}=(m_u+m_d)/2$. Comparing this expression with the corresponding
one obtained from the 1-loop ChPT results,
\begin{eqnarray}
\left( g_\eta -g_\pi\right)_{\rm ChPT} =&&\theta_0F^2\left( {(m_\pi^2
-{\bar m}_K^2)\over 8\pi^2F^2}\log ({\bar m}_K^2/\mu^2)-{B_0(m_s-{\hat m})
\over 8\pi^2F^2}\right. \nonumber\\
&&\left. \qquad\qquad
+{32B_0(m_s-{\hat m})\over 3F^2}L_5^r(\mu^2) +\cdots \right),\label{chptch}
\end{eqnarray}
one sees that the sum-rule expression is lacking both the leading analytic and
leading non-analytic terms in its chiral expansion\cite{km95}, and hence is
incorrect.  Moreover, the numerical consequences of this are significant: the
sum-rule value for the slope is more than an order of magnitude smaller than
that given by ChPT.

\section{Corrected Version of the Sum-Rule Analysis}

The key to understanding the origin of the problem with the CHM sum-rule
analysis lies in Eq.~(\ref{dispreln}).  This relation follows from general
properties of analyticity and unitarity under two assumptions:  (a) that the
singularities of $\Pi_P(q^2)$ consist solely of those associated with physical
intermediate states and (b) that $\Pi_P(q^2)$ converges sufficiently fast that
no subtractions are required.  The latter assumption is explicitly verified by
the known asymptotic behavior of $\Pi_1(q^2)$ and $\Pi_2(q^2)$ in QCD.  The
former, however, is more subtle, since there can also be singularities of
purely kinematic origin. In the case at hand, Eq.~(\ref{pi2chpt}) shows
explicitly that $\Pi_2(q^2)$ has a kinematic pole at $q^2=0$. As a
consequence, the correct version of the dispersion relation
Eq.~(\ref{dispreln}) must include the contribution of this kinematic pole to
the underlying contour integral. Another way of saying this is that it is
$q^2\Pi_2(q^2)$ which satisfies a dispersion relation without kinematic pole
terms.  The dispersion relation for this function, however, requires one
subtraction in order to converge.  The resulting subtraction constant gives
rise to the kinematic pole term of $\Pi_2(q^2)$.  Its value is calculable in
ChPT, and turns out to correspond precisely to the contact contributions given
in Eq.~(\ref{pi2chpt}).

Bearing this in mind, it is straightforward to write down the corrected
dispersion relation for $\Pi_P(q^2)$,
\begin{equation}
\Pi_P(q^2)=-{\frac{1}{q^2}}{\frac{B_0(m_d-m_u)}{\sqrt{3}}}
\Biggl[ {\frac{3}{32\pi^2}}\left(\log (m_K^2/\mu^2)+1\right) -8L_5^r(\mu^2)
\Biggr]+\int {\frac{1}{\pi}}{\frac{{\rm Im}\, \Pi_P(s)}
{s-q^2}}\, ds\ ,
\label{newdispreln}
\end{equation}
where ${\rm Im}\, \Pi_P(s)$ includes only the spectral strength associated with
pseudoscalar states. The corresponding Borel-transformed sum rule is then
\begin{eqnarray}
{\frac{C_2}{2\sqrt{3} M^2}}& =& \Biggl[ \left(
-{\frac{B_0(m_d-m_u)}{\sqrt{3}}}\right)\left(
{\frac{3}{32\pi^2}}\left( \log (m_K^2/\mu^2)+1\right) -8L_5^r\right)
+g_\pi
\ e^{-{m^2_\pi /M^2}} 
\nonumber \\
&&\quad - g_\eta \ e^{-{m^2_\eta / M^2}}
+g_{\eta^\prime}\ e^{-{m^2_{\eta^\prime}/M^2}} 
+g_{\pi^\prime}\ e^{-{m^2_{\pi^\prime}/M^2}} 
+ \cdots\Biggr]\ .
\label{newsr}
\end{eqnarray}

As one might expect, the inclusion of the kinematic-pole contribution
cures the problem of the incorrect chiral behavior of $g_\eta -g_\pi$.
To see this, consider the ${\cal O}(M^0)$ terms of Eq.~(\ref{newsr}).
Bearing in mind that $g_{\eta^\prime}$, $g_{\pi^\prime}$, \dots\ are all
of order ${\cal O}(m_q^2)$\cite{svz2}, one has
\begin{equation}
0= \left( -{\frac{B_0(m_d-m_u)}{\sqrt{3}}}\right)\left(
{\frac{3}{32\pi^2}}\left( \log (m_K^2/\mu^2)+1\right) -8L_5^r(\mu^2)\right)
+g_\pi -g_\eta +{\cal O}(m_q^2)\ ,
\label{modsr}
\end{equation}
where the first term on the RHS results from the kinematic pole in
Eq.~(\ref{newdispreln}).  Without this term, one gets $g_\eta -g_\pi ={\cal
O}(m_q^2)$, as found by CHM.  In contrast, using the corrected sum rule, one
finds that Eq.~(\ref{modsr}) is simply an alternate form of Eq.~(\ref{chptch}),
as required.

To clarify the physical content of the remaining pieces of the sum rule,
Eq.~(\ref{newsr}), it is useful to note the chiral order of various quantities
appearing therein. In particular, the chiral expansions of $g_\pi$, $g_\eta$,
$m_{\eta^\prime}^2$ and $m_{\pi^\prime}^2$ start at order ${\cal O}(p^0)$,
$m_\pi^2$, $m_\eta^2$ and $C_2$ at ${\cal O}(p^2)$, and (as already noted
above) $g_{\eta^\prime}$, $g_{\pi^\prime}$ at ${\cal O}(m_q^2)={\cal O}(p^4)$.
After the cancellation embodied in Eq.~(\ref{modsr}), the only ${\cal O}(p^2)$
terms remaining in Eq.~(\ref{newsr}) are those in $C_2$ and $-g_\pi
m_\pi^2+g_\eta m_\eta^2$.  Using the leading-order expressions $\langle\bar
uu\rangle=\langle\bar dd\rangle=-B_0 F^2$ and $g_\pi =g_\eta =\theta_0 F^2$,
it is straightforward to show that the ${\cal O}(p^2)$ terms on both sides of
the sum rule also match properly. To this order, the matching is just an
isospin-breaking version of the Gell-Mann--Oakes--Renner relation.

The information obtained in the previous paragraph is all that we can extract
 from Eq.~(\ref{newsr}) in its present form. This is because terms of ${\cal
O}(p^4)$ have not been included on the OPE side of the sum rule. If one wishes
to use the sum rule to obtain information about anything beyond the leading and
next-to-leading order behavior of $g_\pi$ and $g_\eta$, one must, therefore,
restore the ${\cal O}(m_q^2)$ terms to the OPE.  This is easily accomplished
starting from the expression for the corresponding terms in the OPE of the
analogous isospin-conserving correlator, as given in Ref.~\cite{BNdRY}. The
result is
\begin{eqnarray}
\left[ \Pi_P(q^2)\right]_{OPE}=-{\frac{1}{\sqrt{3} Q^2}}&&\Biggl[
{\frac{3(m_d^2-m_u^2)}{8\pi^2}}\log \left( {\frac{Q^2}{\mu^2}}\right)
+{\frac{m_d\langle\bar d d\rangle-m_u\langle\bar u u\rangle}{Q^2}}\nonumber \\
&&\qquad +
{\frac{(m_d^2-m_u^2)}{\pi Q^4}}\langle\alpha_s G^2\rangle\Biggr]\ ,
\label{om2sr}
\end{eqnarray}
where we have kept terms only up to dimension $4$ and written down the
coefficient functions only to leading order in $\alpha_s$. Substituting the
expression $\left[ \Pi_P(q^2)\right]_{OPE}$ into the LHS of
Eq.~(\ref{newdispreln}) and Borel transforming, we obtain an improved
version of the corrected CHM sum rule, Eq.~(\ref{newsr}).  To facilitate
subsequent analysis, it is convenient to multiply both expressions for the
correlator by $Q^2$ before Borel transforming (thereby eliminating the
contribution of the kinematic pole).  We also follow standard
practice and introduce a continuum threshold parameter, $s_0$, representing
the point beyond which the hadronic spectral function is modelled by its
perturbative QCD counterpart. The contribution corresponding to the integral
over that portion of the phenomenological spectral function can then be moved
to the OPE side of the sum rule. The result of these manipulations is the sum
rule,
\begin{eqnarray}
g_\pi &&m_\pi^2\ e^{-m_\pi^2/M^2}-g_\eta m_\eta^2\ e^{-m_\eta^2/M^2}
+\sum_{P\not= \pi ,\eta}g_Pm_P^2\ e^{-m_P^2/M^2}\nonumber\\
&&= {\frac{1}{\sqrt 3}}\Biggl[
{\frac{3(m_d^2-m_u^2)}{8\pi^2}}M^2\left( e^{-s_0/M^2}-1\right)
+ \left( m_d\langle\bar d d\rangle -m_u\langle\bar u u\rangle\right)
-{\frac{(m_d^2-m_u^2)}{\pi M^2}}\langle\alpha_s G^2\rangle\Biggr]\ ,
\label{finalsr}
\end{eqnarray}
where the sum on the LHS now runs over pseudoscalar resonances with
squared masses less than $s_0$.

The chiral expansion of the sum rule, Eq.~(\ref{finalsr}), contains terms of
order ${\cal O}(p^2)$ and higher, together with the usual chiral logs, which
start at order ${\cal O}(p^4\ln p)$. The ${\cal O}(p^2)$ terms are the same as
those in Eq.~(\ref{newsr}) and so it is easy to see the sum rule is consistent
to this order. Since only the light-quark condensate and the quantities
$g_\pi$, $g_\eta$, $m_\pi^2$ and $m_\eta^2$ contain leading chiral logs, these
contributions must also cancel in Eq.~(\ref{finalsr}) (as verified below).
Finally, the expansion of $g_\pi m_\pi^2\ e^{-m_\pi^2/M^2} -g_\eta m_\eta^2\
e^{-m_\eta^2/M^2}$ to order ${\cal O}(p^4)$ can be found from known 1-loop
expansions, and that for $ m_d\langle\bar d d\rangle -m_u\langle\bar u
u\rangle$ can be obtained from a straightforward 1-loop calculation. With
these results, we may employ this sum rule to obtain a relation describing the
leading chiral behavior (${\cal O}(m_q^2)={\cal O}(p^4)$) of the
isospin-breaking parameters $g_{\eta^\prime}$, $g_{\pi^\prime}$, \dots\ for
the heavy pseudoscalar mesons.  No further information can be extracted from
Eq.~(\ref{finalsr}) without 2-loop ChPT calculations as input.

To verify the cancellation of the chiral logs, and to obtain the promised sum
rule for the leading chiral behavior of $g_{\eta^\prime}$, $g_{\pi^\prime}$,
\dots, we expand the $\pi$, $\eta$ and condensate terms in
Eq.~(\ref{finalsr}) to order ${\cal O}(p^4)$.  To do so for the $\pi$ and
$\eta$ terms appearing on the LHS requires only the 1-loop expressions for
$f_\pi$, $f_\eta$, $\epsilon_1$ and $\epsilon_2$ given by Gasser and
Leutwyler\cite{gl85}.  The results are
\begin{eqnarray}
g_\pi m_\pi^2 -g_\eta m_\eta^2 &=& -\theta_0 F^2\Biggl[
{\frac{4}{3}}B_0(m_s-\hat{m})-B_0\left( {\frac{
9(m_s-2\hat{m})\ell_\pi +6m_s\ell_K +(m_s+2\hat{m})\ell_\eta }
{72\pi^2F^2}}\right)\nonumber \\
&&\qquad\qquad 
+{\frac{128B_0^2(m_s-\hat{m})}{3F^2}}\Bigl[ (m_s+2\hat{m})L_6^r(\mu^2)
-2(m_s-\hat{m})L_7^r+2\hat{m}L_8^r(\mu^2)\Bigr] 
\nonumber \\
&&\qquad\qquad +{\frac{B_0^2\hat{m}^2}{6\pi^2F^2}}\log 
\left( m_K^2/\mu^2\right)-{\frac{B_0^2\hat{m}(m_s-\hat{m})}{12\pi^2F^2}}
\Biggr]
\label{o41}
\end{eqnarray}
and
\begin{equation}
{\frac{1}{M^2}}\left( -g_\pi m_\pi^4 +g_\eta m_\eta^4\right) =
{\frac{16\theta_0 F^2}{9M^2}}B_0^2(m_s-\hat{m})^2\ ,
\label{o42}
\end{equation}
where $\ell_P=m_P^2\log \left( m_P^2/\mu^2\right)$ and all other
notation is as in Gasser and Leutwyler\cite{gl85}. For the condensate
contributions on the RHS, we require the expressions for $\langle\bar d
d\rangle$ and $\langle\bar u u\rangle$ valid to order ${\cal O}(m_d-m_u)$.
These are easily obtained, and can be written
\begin{eqnarray}
\langle\bar u u\rangle&=&\langle\bar u u\rangle_I + \delta \nonumber \\
\langle\bar d d\rangle&=&\langle\bar u u\rangle_I - \delta
\label{o43}
\end{eqnarray}
where $\langle\bar u u\rangle_I$ is the 1-loop expression for the condensate
in the isospin-symmetric limit, also to be found in Ref.\cite{gl85}, and
\begin{equation}
\delta = (m_d-m_u)\Biggl[ {\frac{\ell_\eta -\ell_\pi}{64\pi^2F^2(m_s-\hat{m})}}
+{\frac{B_0\left( 1+\log \left( m_K^2/\mu^2\right)\right)}{32\pi^2F^2}}
-{\frac{4B_0}{F^2}}\left( 2L_8^r(\mu^2) +H_2^r(\mu^2)\right)\Biggr]\ .
\label{o44}
\end{equation}
 From Eqs.~(\ref{o43}) and (\ref{o44}) it follows that, to order ${\cal
O}(p^4)$,
\begin{eqnarray}
{\frac{\left[ m_d\langle\bar d d\rangle-m_u\langle\bar u
u\rangle\right]}{\sqrt 3}}&=& -\theta_0 F^2\Biggl[
{\frac{4}{3}}B_0(m_s-\hat{m})
-{\frac{B_0^2\hat{m}(m_s-\hat{m})}{12\pi^2F^2}}
+{\frac{B_0^2\hat{m}^2}{6\pi^2F^2}}\log \left({\frac{ m_K^2}{\mu^2}}\right)
\nonumber \\
&&-B_0\left( {\frac
{9(m_s-2\hat{m})\ell_\pi +6m_s\ell_K +(m_s+2\hat{m})\ell_\eta}{72\pi^2F^2}} 
\right) \nonumber \\
&&+{\frac{64B_0^2(m_s-\hat{m})}{3F^2}}
\Bigl[ 2(m_s+2\hat{m})L_6^r(\mu^2)
+2\hat{m}L_8^r(\mu^2) + \hat{m}H_2^r(\mu^2) \Bigr] \Biggr].
\label{cond}
\end{eqnarray}

To obtain a sum rule for the leading chiral behavior of the higher pseudoscalar
resonances, we make use of Eqs.~(\ref{o41}) and (\ref{o42}) to replace the
leading terms of the $\pi$ and $\eta$ contributions in Eq.~(\ref{finalsr}). 
The terms of higher order in $m_\pi$ and $m_\eta$ may be neglected since they
are at least of order ${\cal O}(p^6)$ in the chiral expansion, and they are
numerically small for the Borel masses of interest. Finally, inserting the
chiral expansion of the quark condensates, Eq.~(\ref{cond}), into this sum
rule, we get
\begin{eqnarray}
\sum_{P\not= \pi ,\eta}\, g_Pm_P^2\ e^{-m_P^2/M^2}= &&{\frac{\sqrt 3 M^2}
{8\pi^2}}\left( e^{-s_0/M^2}-1\right) (m_d^2-m_u^2)
+{\frac{(m_d^2-m_u^2)}{8\sqrt{3}\pi M^2}}\langle\alpha_s G^2\rangle
\nonumber \\
&&-{\frac{64B_0^2(m_s-\hat{m})\theta_0}{3}}
\left[ 4(m_s-\hat{m})L_7^r -2\hat{m}L_8^r(\mu^2)+\hat{m}H_2^r(\mu^2)\right] 
\nonumber\\
&&-{\frac{16}{9M^2}}\theta_0F^2B_0^2(m_s-\hat{m})^2\ .
\label{ressr}
\end{eqnarray}
Note that all of the chiral logarithms have cancelled, leaving only terms that
start at order ${\cal O}(p^4)$.

It is worth noting that the term involving the chiral LEC's makes a
numerically significant contribution to the sum rule and is dominated by
$L_7^r$. Moreover, phenomenological treatments that use resonance exchanges to
generate the LEC's in the effective Lagrangian of ChPT\cite{EGPdR,Detal} show
that $L_7^r$ receives contributions only from flavor-singlet pseudoscalar
states. Hence it already follows, without any more detailed analysis, that the
sum rule implies the existence of significant isospin-breaking $\eta^\prime$
coupling. Before presenting the results of our analysis for the
isospin-breaking parameters, $g_P$, it is also worth stressing a number of
features of the sum rule, Eq.~(\ref{ressr}), which imply that, once $L_7^r$ is
fixed, these results for at least the $\eta^\prime$ coupling should be quite
reliable.

The renormalized LEC's $L_7^r$, $L_8^r$ have been determined
phenomenologically and are reasonably well-known (see, for example,
Refs.~\cite{bkm95,ecker95} for recently updated values).  The remaining LEC,
$H_2^r$, can be related, for example, to the isospin-breaking condensate ratio
$\gamma\equiv [\langle\bar d d\rangle-\langle\bar u u\rangle]/\langle\bar u
u\rangle$. This ratio has been estimated in a number of sum-rule
analyses\cite{narison87,ddr87,narison89,narisonbk,adi93,ei93}. Using any of
the values of $\gamma$ obtained in these treatments to estimate $H_2^r$, the
resulting values are such that the LEC combination in Eq.~(\ref{ressr}) is
dominated by the $L_7^r$ term. In particular, the uncertainty in the LEC
combination associated with the sum of the errors on the phenomenological
determinations of $L_8^r$ and $H_2^r$ (where the latter error is taken to
correspond to the entire range of values cited above) is an order of magnitude
smaller than that associated with the error on the existing phenomenological
determination of $L_7^r$.  We may, therefore, ignore the effect of the
uncertainties in the values of $L_8^r$ and $H_2^r$. This feature of the
analysis results from the fact that the coefficients of $L_8^r$ and $H_2^r$ are
suppressed by a factor of $(m_s-\hat{m})/\hat{m}\sim 23$\cite{leutwyler96}
relative to that of $L_7^r$. The uncertainty in the ratio $m_s/\hat{m}$ which
enters this suppression is, of course, also completely negligible. Note that
we do not require an explicit input value for $m_s$ since, to the order
considered in the chiral expansion, we may take
\begin{equation}
B_0^2(m_s-\hat{m})^2=(m_K^2-m_\pi^2)^2\ .
\end{equation}

On the phenomenological side of the sum rule, we expect contributions from all
of the higher pseudoscalar resonances, $\eta^\prime (958)$, $\eta (1295)$,
$\pi^\prime (1300)$, $\eta (1440)$, $\pi^\prime (1800)$, \dots. The $\pi^\prime
(1300)$ is relatively broad ($\Gamma = 325 \ {\rm MeV}$\cite{pdg96}) and spans
the region between the $\eta (1295)$ and the $\eta (1440)$.  Therefore,
without keeping terms of yet higher dimension in the OPE, we have too little
information in the sum rule to both adequately parametrize the spectral
function in the region between $\sim 1300$ and $\sim 1450$ MeV and at the same
time to use the sum rule to extract the values of all such parameters.  Hence
we concentrate on the extraction of $g_{\eta^\prime}$, parametrizing the $\eta
(1295)$, $\pi^\prime (1300)$, $\eta (1440)$ region in terms of a single
effective contribution of zero width located at around $1375$ MeV.  By varying
the position of this contribution between $1300$ and $1450$ MeV, we have
verified that the extracted value of $g_{\eta^\prime}$ is not sensitive to this
approximation, varying by $\sim\pm 6\%$ over this range. This is a factor of 6
smaller than the variation induced by the uncertainty in the input value of
$L_7^r$, which we discuss in more detail below.

The effective strength parameter describing the $\eta (1295)$, $\pi^\prime
(1300)$ and $\eta (1440)$ region (which we denote by $g_{\pi^\prime}$ in what
follows) is, of course, much more sensitive to the assumed position of this
strength. The corresponding uncertainty in the extraction of $g_{\pi^\prime}$
is $\sim 15\%$, which is significant, although still much less than the $\sim
60\%$ associated with $L_7^r$. The stability of the determination of
$g_{\eta^\prime}$ is attributable, to a large extent, to the fact that the
residual term proportional to $L_7^r$ provides the major contribution to the
sum rule; as already noted above, $L_7^r$ is known to receive contributions
only from flavor-singlet states, of which the $\eta^\prime$ is nearest and
hence should provide the dominant contribution. This feature of the sum rule
is also responsible for the greater sensitivity of $g_{\pi^\prime}$ to the
input value chosen for the location of the effective strength describing the
$\eta (1295)$, $\pi^\prime (1300)$, and $\eta (1440)$ region:  the combined
effective contribution to the sum rule is small relative to the dominant
$\eta^\prime$ term and the extracted value can therefore depend sensitively on
the assumed separation from the $\eta^\prime$ peak.

Having employed information from ChPT to fix the low-lying $\pi$ and $\eta$
contributions to the original sum rule, and explicitly modelled the
contributions up to $1.44$ GeV, we note that there is now a significant gap to
the next resonance contribution at $1.8$ GeV. We therefore expect that Borel
masses of order $1-1.5$ GeV will suppress the contributions of higher
resonance on the phenomenological side of the sum rule. 

On the OPE side it turns out that the situation is also rather favorable.
First, the gluon condensate term turns out to be numerically very small
compared to the dominant $L_7^r$ contribution.  Indeed, if we take for
definiteness the value for this condensate advocated in Ref.~\cite{leinweber}
(which is similar to that employed, for example, in Ref.~\cite{bpr}),
\begin{equation}
\langle {\frac{\alpha_s}{\pi}}G^2\rangle = 0.03\pm 0.015\ {\rm GeV}^4\ ,
\label{gcond}
\end{equation}
which includes rather conservative errors, then we find that this uncertainty
corresponds to $<0.3\%$ variations in $g_{\eta^\prime}$ and $g_{\pi^\prime}$.

The perturbative contribution (the first term on the RHS of Eq.~(\ref{ressr}))
is similarly small.  This is fortunate since recent
analyses\cite{bg96,bgm97,cfnp97} suggest that conventional sum-rule
determinations of the light current quark masses\cite{bpr,jm95,cps} may have
overestimated these masses by as much as a factor of $2$.  For the central
value of $L_7^r$, allowing $m_u+m_d$ to vary between the conventional value,
$12$ MeV\cite{bpr} and $6$ MeV produces a variation of only $2.5\%$ in 
$g_{\eta^\prime}$ and $g_{\pi^\prime}$.  Such an uncertainty is again much
smaller than that arising  from the errors on $L_7^r$, and hence can be
neglected. The smallness of this perturbative contribution also implies that
the analysis should be rather insensitive to the continuum threshold
parameter, $s_0$. We expect that this should lie somewhere in the vicinity of
the onset of the $\pi^\prime (1800)$ resonance. In our analysis, we find, for
example, that varying $s_0$ by $\pm 1$ GeV$^2$ about a central value $s_0=3\
{\rm GeV}^2$ produces variations of $<1\%$ in $g_{\eta^\prime}$ and
$g_{\pi^\prime}$.

 From the above discussion, we see that the RHS of the sum rule in
Eq.~(\ref{ressr}) is dominated by the terms that are directly calculable using
ChPT.  The first of these, involving the ${\cal O}(p^4)$ LEC's, is the piece
of the quark condensate term from the OPE that remains after cancellation
against $\pi$ and $\eta$ contributions from the phenomenological side of
the sum rule.  The second consists of the remaining ${\cal O}(m_q^2)$ $\pi$
and $\eta$ contributions from the phenomenological side.  Numerically it is
more than a factor of $2$ smaller than the LEC term, for $M>1 \ {\rm GeV}^2$,
and of the same sign. The major uncertainty in the values of these terms is
that arising from the phenomenological determination of the (scale-independent)
LEC\cite{bkm95,ecker95},
\begin{equation}
L_7^r=(-0.4\pm 0.15)\times 10^{-3}\ .
\label{l7val}
\end{equation}

For completeness we list below the remaining input values (apart from 
well-determined meson masses):
\begin{eqnarray}
m_u+m_d&=& 9\ {\rm MeV}\nonumber \\
\langle {\frac{\alpha_s}{\pi}}G^2\rangle &=& 0.03\ {\rm GeV}^4\nonumber \\
s_0&=& 3.0\ {\rm GeV}\nonumber \\
L_8^r(m_\rho^2) &=& 0.9\times 10^{-3}\nonumber \\
H_2^r(m_\rho^2) &=& -7.5\times 10^{-4}\nonumber \\
m_{\pi^\prime}&=& 1.375\pm 0.075\ {\rm GeV}\nonumber \\
m_s/\hat{m}&=& 24.4\ ,\nonumber \\
r={\frac {m_d -m_u}{m_d+m_u}}&=&0.3\pm 0.05\ ,
\label{input}
\end{eqnarray}
where by $m_{\pi^\prime}$ we mean the location of the effective strength for
the $\eta (1295)$, $\pi^\prime (1300)$, $\eta (1440)$ region, as discussed
above. In most cases we have not shown the corresponding uncertainties, since,
as already noted, the variations in the results associated with them are small.
Apart from $L_7^r$, the largest uncertainty is that associated with the choice
$m_{\pi^\prime}$, which parametrizes the strength lying above the $\eta'$.

Also significant is the uncertainty associated with the isospin-breaking
mass ratio, $r$\cite{leutwyler96}.  The quoted range covers a wide range of
possibilities for the degree of breaking of Dashen's theorem\cite{dashen} for
the electromagnetic contribution to the kaon mass splitting.  The recent
results of Refs.~\cite{dp96,det96,bp96} would appear to confirm a larger value
for the breaking, as suggested by earlier analyses\cite{mk90,dh,bijnens}, and
hence larger values of $r$ in the quoted range, with a somewhat smaller
resulting error.  Since the subject is not yet fully resolved (see
Ref.~\cite{dp96} for a detailed list of recent work on the subject, including
some work advocating smaller violations of Dashen's theorem\cite{bu96}), we
have refrained from attempting to make a revised estimate for the input
central value and error on $r$.  In any case, every term on the RHS of
Eq.~(\ref{ressr}) contains one factor of $m_d-m_u$, so that this uncertainty
enters only into the overall normalization of the final results. It does not,
therefore, affect the stability analysis of the sum rule, and it can be
removed by quoting results in the form $g_P/\theta_0 F^2$.

For a given set of values for the input parameters $L_7^r$ and $m_{\pi'}$, we
look for values of $g_{\eta'}$ and $g_{\pi'}$ that bring the two sides of the
sum rule into agreement over a range of Borel mass values. A convenient way to
do this is to use the sum rule, Eq.~(\ref{ressr}), and its derivative with
respect to $M$, at a fixed value of the Borel mass, as simultaneous linear
equations for $g_{\eta'}$ and $g_{\pi'}$. If a region is found where the
results of this procedure are independent of $M$, then this indicates the
existence of a stability window where the two sides of the sum rule match. In
Fig.~1 we show some typical results for a case where we obtain good stability,
$L_7^r=-0.34\times 10^{-3}$ and $m_{\pi'}=1375$ MeV, with
$g_{\eta'}=2.88\times 10^{-5}$ and $g_{\pi'}=-5.57\times 10^{-6}$.  The
two curves are essentially indistinguishable, except at the very lower
end of Borel masses displayed.

As $|L_7^r|$ is decreased, the stability window moves to larger values of $M$.
In this region, the perturbatively modelled continuum becomes increasingly
important in the spectral representation of the correlator and so the sum rule
becomes unreliable for the determination of resonance properties. In contrast,
as $|L_7^r|$ is increased the stability window moves to smaller values of $M$
and also becomes very much narrower. In fact, for values of $|L_7^r|$ that are
larger than about $0.48\times 10^{-3}$ we are unable to find a stable matching
between the two sides of the sum rule. This occurs before the window reaches 
sufficiently small values of $M$ that the convergence of the OPE becomes 
questionable. We are thus able to use the sum rule to make a somewhat improved
determination of the LEC $L_7^r$, reducing by about a factor of 
$2$ the distance to the upper bound
on its magnitude compared to the ChPT result,
Eq.~(\ref{l7val})\cite{bkm95,ecker95}.

For values of $L_7^r$ in the range $-0.25\times 10^{-3}$ to $-0.48\times
10^{-3}$, we obtain
\begin{eqnarray}
g_{\eta^\prime}/\theta_0 F^2&=& 0.42\pm 0.15\nonumber \\
g_{\pi^\prime}/\theta_0 F^2&=& -0.13\pm 0.07\ .
\label{gscvals}
\end{eqnarray}
The dependence on $r$ has been scaled out of these results, as discussed above,
and so the dominant uncertainties quoted in Eqs.~(\ref{gscvals}) are those
associated with the range of values for $L_7^r$. Allowing for the uncertainty
in $r$ taken from\cite{{leutwyler96}}, our values for the isospin-breaking
parameters are
\begin{eqnarray}
g_{\eta^\prime}&=& (3.6 \pm 1.9)\times 10^{-5}\ {\rm GeV^2} \nonumber \\
g_{\pi^\prime} &=& (-1.1 \pm 0.8)\times 10^{-5}\ {\rm GeV^2}\ .
\label{gvals}
\end{eqnarray}

\section{Summary}

In this paper, we have revisited the sum-rule treatment for the
isospin-breaking axial correlator, correcting the error in a previous
treatment which led to the incorrect chiral behavior of the slope parameter
$g_\eta -g_\pi$.  Including the kinematic pole omitted from the previous
treatment restores the correct chiral behavior of the correlator. We have then
used the explicit evaluation of the $\pi$ and $\eta$ contributions to the
correlator at next-to-leading order in ChPT to obtain a rather well-behaved
sum rule for the leading chiral behavior of the isospin-breaking parameters,
$g_P$, of the higher pseudoscalar resonances.  This sum rule has been analyzed
and shown to provide a rather reliable estimate for $g_{\eta^\prime}$, once
one has fixed the chiral LEC, $L_7^r$.  The requirement of the stability of
this sum rule is shown, moreover, to provide a somewhat improved determination
this LEC by reducing the upper bound on its magnitude.

\acknowledgements
We are grateful for the hospitality of the Special Research Center for the
Subatomic Structure of Matter at the University of Adelaide where much of this
work was performed. We also thank J. McGovern for a critical reading of the
manuscript. KM acknowledges the continuing financial support of the Natural
Sciences and Engineering Research Council of Canada, and MCB that of the UK
Engineering and Physical Sciences Reseacrh Council.
\begin{figure}[htb]
  \centering{\
     \epsfig{angle=0,figure=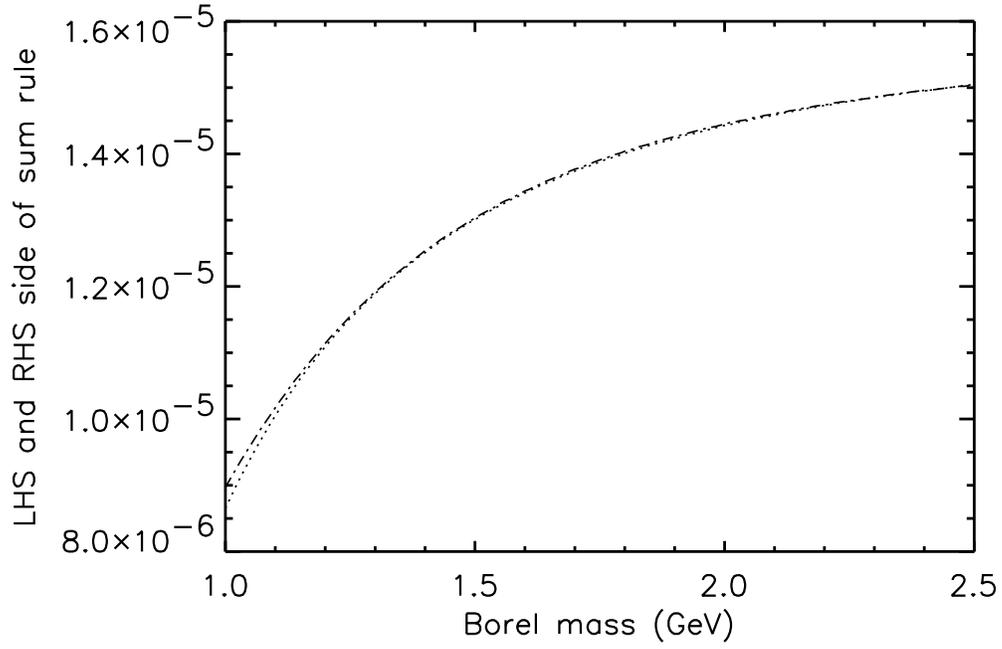,width=0.8\linewidth}
               }
\parbox{130mm}{\caption
{The OPE versus the phenomenological side of the improved sum rule as
a function of the Borel mass, $M$.  The
dotted line is the OPE side, the dash-dotted line the phenomenological
side in units of GeV$^2$.}
\label{figmu} }
\end{figure}

\end{document}